\documentclass[twocolumn,10pt,floatfix,nofootinbib]{revtex4-1}

\usepackage[utf8]{inputenc}
\usepackage[T1]{fontenc}
\usepackage[ngerman,english]{babel}

\usepackage{graphicx}   
\usepackage{booktabs}
\usepackage{amssymb}   
\usepackage{amsmath}   
\usepackage{hyperref}
\usepackage{color}
\usepackage{lineno}

\begin{document}

\title{Image Translation for Medical Image Generation: Ischemic Stroke Lesion Segmentation}

\author{Moritz Platscher} %
\thanks{Corresponding author: \href{mailto:platscher@biomed.ee.ethz.ch}{platscher@biomed.ee.ethz.ch}}
\author{Jonathan Zopes} %
\author{Christian Federau}
\thanks{MP is supported by the University of Zurich under the Grant no.~FK-20-052. JZ acknowledges support from the SPARK Grant CRSK-3\_190697 of the Swiss National Science Foundation. CF is supported by an Ambizione Fellowship grant (PZ00P3\_173952) of the Swiss National Science Foundation. Calculations for this project were performed in part using support from Google and on a Titan Xp donated by the NVIDIA Corporation} 
\thanks{Code available at \url{www.github.com/MoPl90/image_translation}}
\address{Institute for Biomedical Engineering, University and ETH Zürich, Gloriastrasse 35, 8092 Zürich, Switzerland}

\begin{abstract}

Deep learning based disease detection and segmentation algorithms promise to improve many clinical processes. However, such algorithms require vast amounts of annotated training data, which are typically not available in the medical context due to data privacy, legal obstructions, and non-uniform data acquisition protocols. Synthetic databases with annotated pathologies could provide the required amounts of training data. We demonstrate with the example of ischemic stroke that an improvement in lesion segmentation is feasible using deep learning based augmentation. To this end, we train different image-to-image translation models to synthesize magnetic resonance images of brain volumes with and without stroke lesions from semantic segmentation maps. In addition, we train a generative adversarial network to generate synthetic lesion masks. Subsequently, we combine these two components to build a large database of synthetic stroke images. The performance of the various models is evaluated using a U-Net which is trained to segment stroke lesions on a clinical test set. We report a Dice score of $\mathbf{72.8}$\% [$\mathbf{70.8\pm1.0}$\%] for the model with the best performance,  which outperforms the model trained on the clinical images alone $\mathbf{67.3}$\% [$\mathbf{63.2\pm1.9}$\%], and is close to the human inter-reader Dice score of $\mathbf{76.9}$\%. Moreover, we show that for a small database of only 10 or 50 clinical cases, synthetic data augmentation yields significant improvement compared to a setting where no synthetic data is used. To the best of our knowledge, this presents the first comparative analysis of synthetic data augmentation based on image-to-image translation, and first application to ischemic stroke. 

\end{abstract}


\maketitle

\begin{figure*}[t]
    \centering
    \includegraphics[width=.85\textwidth, trim = 0 0 0 3.6cm, clip]{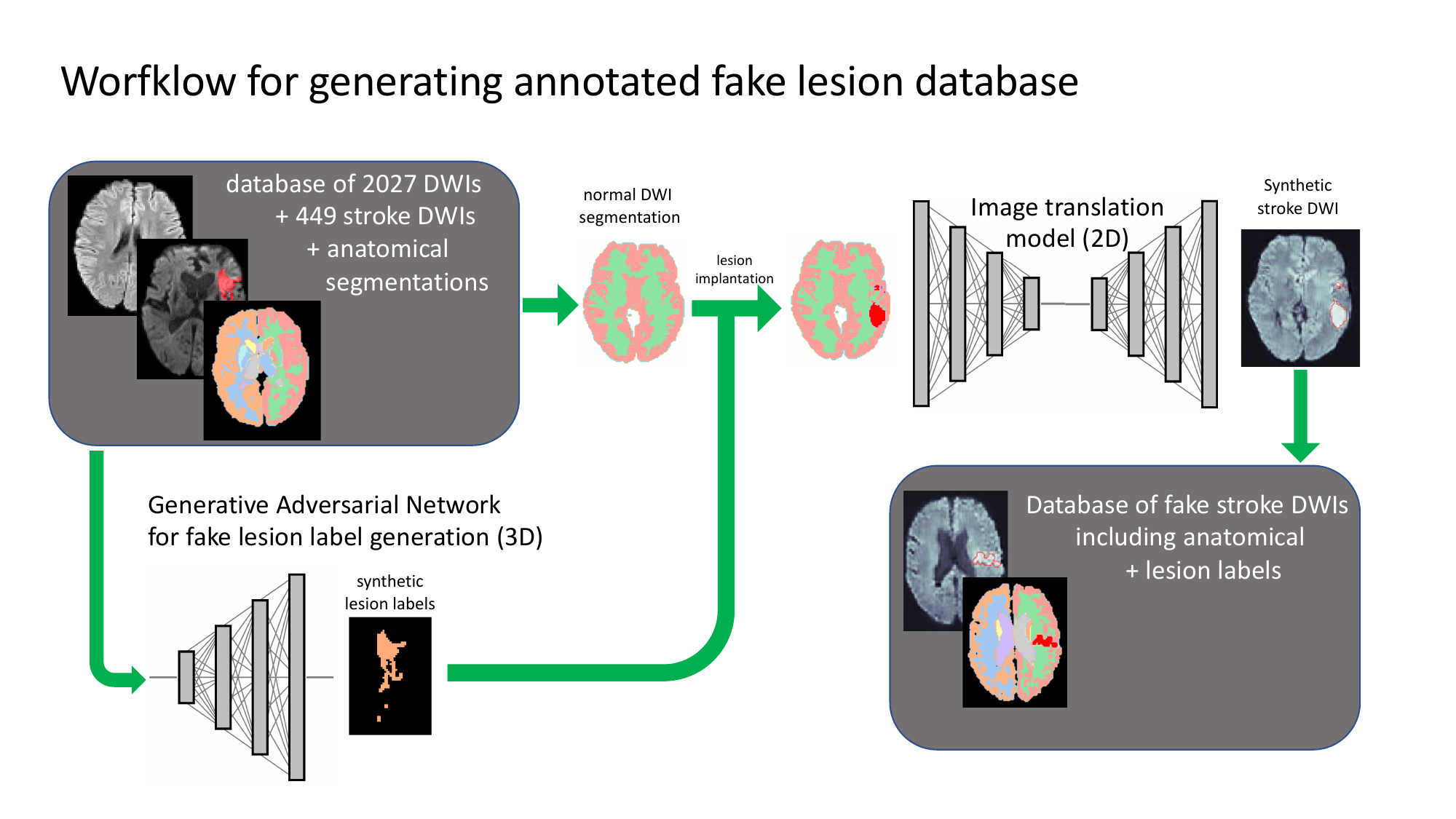}
    \caption{A GAN is trained to produce lesion label maps from a database of 365 manually segmented stroke patients' DWIs. Using the same 365 pathological volumes and another 2027 healthy brain MR scans, various ITMs are trained to synthesize DWIs from anatomical segmentation masks. The trained ITM can then be used to combine fake lesion labels with healthy brain segmentation masks (real or fake) to produce synthetic stroke DWIs with high quality segmentation labels.}
    \label{fig:workflow}
\end{figure*}

\section{Introduction}

Ischemic stroke (IS) is the second leading cause of death worldwide~\cite{WHOcauseofdeath}, and, if not fatal, frequently results in irreversible brain tissue damage and disabilities. IS is caused by occlusion of blood vessels, resulting in restricted or suppressed blood supply to some areas of the brain. Consequently, the success of therapy is closely tied to the time between the onset of symptoms and successful revascularization treatment~\cite{timetotreatment1,timetotreatment2}. Therefore, it is of utmost importance to streamline and optimize the diagnostic process.
While many clinical centers use computed tomography (CT) as a first imaging modality for stroke due to its wider availability, speed, and relative absence of contraindications, the gold standard for the identification and quantification of IS lesions is diffusion-weighted magnetic resonance imaging (DWI), on which the infarcted region induces a hyper-intensity compared to normal tissue, corresponding to zones of relative diffusion restriction. Manual segmentation of these areas is a relatively simple, but tedious task for a trained expert, while automatic segmentation using standard methods is a difficult task, e.g., due to the complex lesion geometries, non-trivial variability in magnetic resonance signal intensities, varying locations, and numerous image artifacts. Machine learning, and deep learning (DL) in particular, holds the potential to robustly automatize such tasks~\cite{AIinMedicine,BERNAL201964,ReviewMedImaging}. 
However, the training of DL models requires vast amounts of labeled data, which are typically not available due to privacy concerns, incompatible data formats, or because the disease is rare. In addition, manual segmentation of medical images is a task that only experienced radiologists can perform, is expensive, error-prone, and  time-consuming. Consequently, large databases of annotated medical images are scarce: For example, the Ischemic Stroke Lesion Segmentation Challenge (ISLES) data set contains  64 IS volumes~\cite{ISLES}. In comparison, the ImageNet database~\cite{ImageNet}, a popular data set for object recognition in the non-medical area, contains 14~million~images.

Synthetic image generation with deep generative models~\cite{Goodfellow2014generative} could bridge the success of DL to medical applications by augmenting available data sets, or replacing them altogether, circumventing the aforementioned concerns related to data security and availability. However, these models, too, require large training data sets in order to obtain good results. Moreover, alongside realistic images, the model must provide lesion labels, which is not straightforwardly implemented in the current deep generative pipelines. 

We attempt to synthesize DWIs of stroke lesions with image-to-image translation models (ITMs), i.e.~DL models that transform images from one domain of training data to another domain of training data: semantic segmentation maps containing anatomical labels of the brain structure and the label of the stroke lesion are translated into brain DWIs. In addition, new lesion labels are generated by a 3D generative adversarial network. We envision that the anatomical labels will guide the network to produce sharp images with tissue contrast comparable to the real DWIs. At the same time, the lesion labels of the training set will provide sufficient information to learn a pathology-specific contrast modification. The ITM should learn to generate hyper-intensities inside the lesion labels, while ignoring any (statistically underrepresented) misclassification, such as hyper-intensities outside lesion labels or normal tissue misclassified as belonging to a lesion by the human reader. This has a welcome side effect: the network will produce DWIs with lesions that match the input region labeled as such with high confidence, thus providing accurate lesion labels.
The synthetic images obtained using such an approach can then be used to augment a clinical database, or, such as here, directly for supervised learning to segment stroke lesions from normal brain tissue.

\paragraph*{Related work}
The proposed approach is similar in spirit, but more general than Ref.~\cite{FederauStroke}, where synthetic IS volumes were generated by fusing real IS lesion contours into healthy brain DWIs by linearly increasing the voxel intensity within the contour to mimic a lesion. This ensures that anatomical boundary conditions are respected and realistic intensity patterns are generated. Part of our proposal is to outsource this task to a generative DL model, which generalizes the allowed transformations given that neural networks are intrinsically non-linear. With their approach the authors achieved an increase in the Dice similarity coefficient (DSC) from 65\% using the clinical data alone to 70\% when including 2,000 synthetic images, and even up to 72\% using 40,000 synthetic DWIs. 
An advantage of this approach is that it generates coherent 3D data; however, it is limited combinatorically since there is only a limited number of real lesion labels available. Moreover, Refs.~\cite{syntheticBrainTumor,8363653,natureTumorSeg} successfully used generative adversarial networks (GANs) to generate synthetic magnetic resonance images (MRIs) to improve brain tumor segmentation. A different approach was used by~\cite{CTtoMRseg}: there the authors use an ITM to obtain DWIs from CT images allowing them to improve the segmentation quality of the ischemic stroke core tissue on CTs augmented with synthetic DWIs. Similarly, Ref.~\cite{revisions1} used CT-to-MR image-translation to improve lung tumor segmentation. Further approaches using style transfer were recently proposed in Ref.~\cite{revisions4}, where image-to-image translation was used to translate healthy into diseased MR images for brain tumor segmentation, and in Ref.~\cite{revisions5} transforming MR into CT images for general treatment planning. Finally, Ref.~\cite{revisions3} introduced an ITM approach that is was used for a variety of medical tasks involving image denoising, motion-correction, and domain adaptation. 

\paragraph*{Contribution}
We provide a new approach to synthesizing both IS lesion labels and DWIs from semantic segmentation maps. While the former task is realized by ITMs, the lesion labels are generated by a 3D GAN. This allows us to build spatially coherent brain volumes including lesion labels. Moreover, this makes our pipeline scalable in the sense that the GAN can be used to produce an unlimited number of synthetic, yet distinct, lesion labels. We thereby demonstrate that DL-based data augmentation is apt to leverage the information contained in limited medical data sets and outperform conventional data augmentation techniques, an idea that was coined by~\cite{antoniou2017data} and further developed for segmentation in~\cite{twoStageAugmentation} in a non-medical context. We quantify this statement by comparing the DSC of a U-Net~\cite{RonnebergerUNet} trained on manually labeled data versus U-Nets trained with the synthetic data, each evaluated on a test set of clinical IS data. We highlight several areas of application for our pipeline and emphasize the utility of the approach, given that the trained ITMs can be freely distributed without any data privacy concerns. A number of studies exist that investigate the potential improvement of downstream clinical tasks, e.g., anatomical segmentation and tumor segmentation; however many discuss only one specific model setup. To the best of our knowledge, this is the first comparative study using ITMs for data augmentation in IS lesion segmentation.

This paper is structured as follows. In Sec.~\ref{sec:methods} we describe the data and models that are used in the image synthesis pipeline and detail how to evaluate the results quantitatively. The results of the analysis are presented and discussed in Sec.~\ref{sec:results}, before concluding in Sec.~\ref{sec:conclusions}. The Appendix contains supplemental information.

\section{Materials and Methods}\label{sec:methods}

We propose an approach for generating annotated DWIs of brains displaying an IS lesion by means of two consecutive generative models, one for generating realistic stroke lesion labels and another to translate brain segmentation masks into DWIs. Anatomical segmentation masks of healthy brains are readily available and require no medical training, since several automatic segmentation algorithms exist. The synthesis of stroke lesion labels by a GAN, on the other hand, is assumed to introduce general transformations, as well as realistic interpolations of latent space representations of available lesion masks. Finally, the use of conditional generative models to generate synthetic images, significantly improves the convergence behavior of the generators compared to unconditional, latent space models. Thereby we construct a minimal toolbox, that can easily be scaled by employing the interpolation capacities of the GAN to synthesize more lesion masks to generate more and more annotated DWIs.
This workflow is shown in Fig.~\ref{fig:workflow}. 
In the following we provide a detailed description of the involved data and methods.

\subsection{Data}\label{sec:data}

A database of 804 DWIs of patients that presented with symptoms of IS was obtained at the University Hospital Basel, for which Institutional Review Board approval was granted. This database (henceforth stroke DB) contains 449 DWI positive cases, i.e.~diagnosed IS lesions (mean patient age $72 \pm 14$~years; 200 left-sided ISs,
193 right-sided ISs, 56 bilateral ISs; 194 female and 255 male). The DWI positive cases were separated randomly for training (365) and testing (74). Additionally, 85 DWI negative samples were included in the test set, which amounts to 159 test samples in total. A separate database of 2027 healthy DWI scans (mean age $38 \pm 24$~years; 1088 female, 939 male) was used for the augmentation pipeline (normal DB). The majority of images in the stroke DB were acquired on a $1.5$\,T scanner (67\% @ $1.5$\,T; 23\% @ 3\,T), while most of the normal DB scans used a 3\,T scanner (68\% @ $1.5$\,T; 22\% @ 3\,T). On average, stroke DB scans were acquired with similar echo and recovery times ($\mathrm{T}_E = 90 \pm 16\, \mathrm{ms}$, $\mathrm{T}_R = 7400 \pm 1300\, \mathrm{ms}$ [normal DB] vs. $\mathrm{T}_E = 100 \pm 2\, \mathrm{ms}$, $\mathrm{T}_R = 7000 \pm 1500\, \mathrm{ms}$ [stroke DB]).

Images were co-registered  to the standard Montreal Neurological Institute atlas~\cite{MNIatlas}, skull-stripped using FSL's `brain extraction tool'~\cite{FSL}, and re-sampled to a standard resolution of $128\times128\times40$ voxels using ANTs~\cite{ANTS}. For better stability during training, the top and bottom four slices were cropped. The voxel intensities were clipped at the $99.5^{\mathrm{th}}$ percentile, while the background was clipped at an absolute voxel intensity of 35. Finally, signal intensities were re-scaled  to the range $[-1,1]$. The anatomical segmentation of brains was obtained using 3D U-Net trained on reference images obtained using FreeSurfer~\cite{freesurfer}, as outlined in Ref.~\cite{Zopes2020}. This reduced the processing time for the entire database to a few hours.

\subsection{Image translation}
While unconditional GANs have been trained to yield state-of-the art results on brain DWIs~\cite{Hirte2020DiffusionWeightedMR}, they do not automatically provide ground truth labels for anatomical structures or pathologies. ITMs, on the other hand, are generative models, which generate samples conditional to an input segmentation map. Thus, a ground truth label is available by construction. Moreover, this approach helps to improve the quality of the generated image by providing boundaries between different instances of a segmentation label. In this section we introduce the ITMs studied in this manuscript, further details can be found in the Appendix.

\begin{figure*}[t]
    \centering
    \includegraphics[width=0.9\textwidth, trim = 0 20mm 0 15mm, clip]{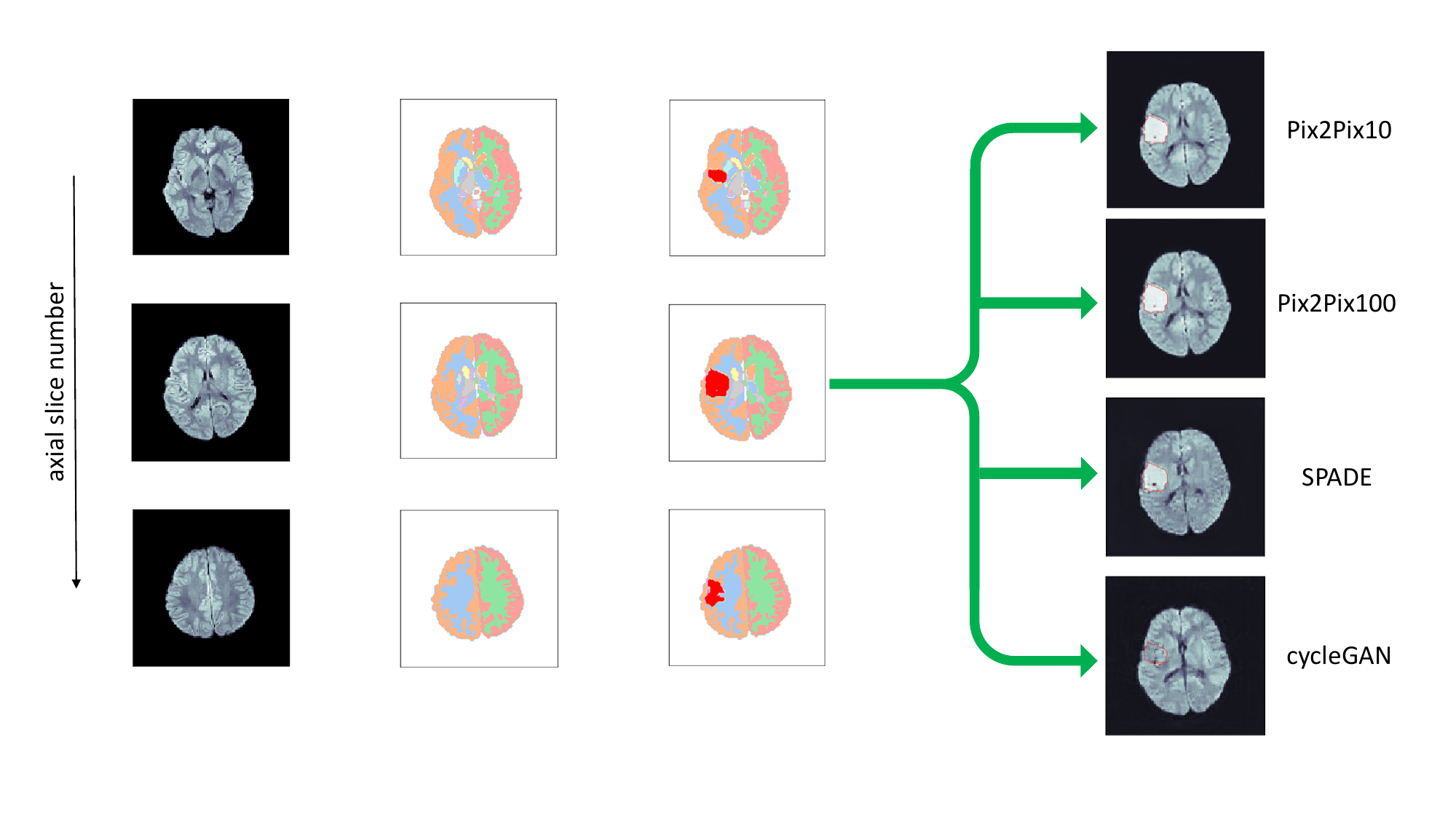}
        \caption{Implantation of a fake lesion and generation of DWIs. \textit{Left}: 3 slices of an original healthy DWI. \textit{Center left}: The segmentation of the original volume. \textit{Center right}: Segmentation map with the implanted lesion label generated by the GAN. \textit{Right}: IS volumes generated from the segmentation maps using different ITMs as indicated.}
    \label{fig:implantation}
\end{figure*}
\subsubsection{Pix2Pix} 
Pix2Pix is a well-studied ITM, which was first proposed by~\cite{pix2pix2017} and further developed by~\cite{pix2pixHD}. Pix2Pix is widely accepted as the method of choice for paired image translation, i.e.~when the images in two domains come in pairs as it is the case for our database. It has previously been applied successfully to medical data, see e.g.~Refs.~\cite{CTtoMRseg,natureTumorSeg, pix2pixMed1,pix2pixMed2,pix2pixMed3,pix2pixMed4}. The original Pix2Pix is based on the U-Net architecture; however, in the high-resolution derivative of the model residual blocks were used~\cite{pix2pixHD,ResNet}. We have trained both architectures but found little difference in quality and therefore used the U-Net based version which converges faster. The loss function for this network is a weighted sum of adversarial loss and $L_1$-norm reconstruction loss,
\begin{equation}\label{eq:ITM_loss}
\begin{aligned}
    \mathcal{L}_\text{Pix2Pix} =& \mathbb{E}_y [\log D(y)] + \mathbb{E}_x [1 - \log D(G(x))] \\ &+ \lambda \| y - G(x) \|_{L_1}\,.
\end{aligned}
\end{equation}
In this equation, $\mathbb{E}_{x/y}$ denotes the expectation value taken w.r.t.~a batch $x$ or $y$ from one of the two domains, $D$ represents the discriminator network which operates on the target domain and $G$ is the generator network.
We consider different values for the reconstruction loss weight $\lambda$, since we have found that a value $\lambda=100$ (Pix2Pix100) yields qualitatively more appealing results than the recommended value $\lambda=10$ (Pix2Pix10). For the discriminator architecture we rely on the PatchGAN~\cite{pix2pix2017,PatchGAN1,PatchGAN2}. In contrast to ordinary discriminators, PatchGAN does not output a single number to characterize if an image is real or fake, but does so for patches of the image, thus allowing a more refined feedback for the generator on smaller scales. The large scale, or low-frequency, features are sufficiently well captured by the $L_1$ loss~\cite{pix2pix2017}.

\subsubsection{cycleGAN}
A method of unpaired image translation is given by cycleGAN~\cite{cycleGAN}. Two generator-discriminator pairs, one for each image domain, allow the model to be trained via a reconstruction loss, translating from domain $A$ to $B$ using the first generator, then back from $B$ to $A$ using the other generator and computing the pixel-wise ($L_1$ or $L_2$ norm) difference between the resulting and the original image. This cycle consistency lends its name to the model and tackles the problem of unpaired data. However, the number of parameters to train is quite vast, reaching 42~million~in our setup. In order to handle this large number of free parameters, the model enforces an identity loss, which penalizes deviations of the generators from an identity mapping.\footnote{Clearly, this identity-loss cannot be used when the input is a multi-channel segmentation map, while the output is a single-channel image. Thus, we do not include an identity loss for training.}
Note that cycleGAN has been applied successfully to paired data in medical image generation~\cite{cycleGANmed1,StanfordCycleGAN}. 

\begin{figure*}[t]
\centering
\includegraphics[width=0.9\textwidth, trim = 0 25mm 0 25mm, clip]{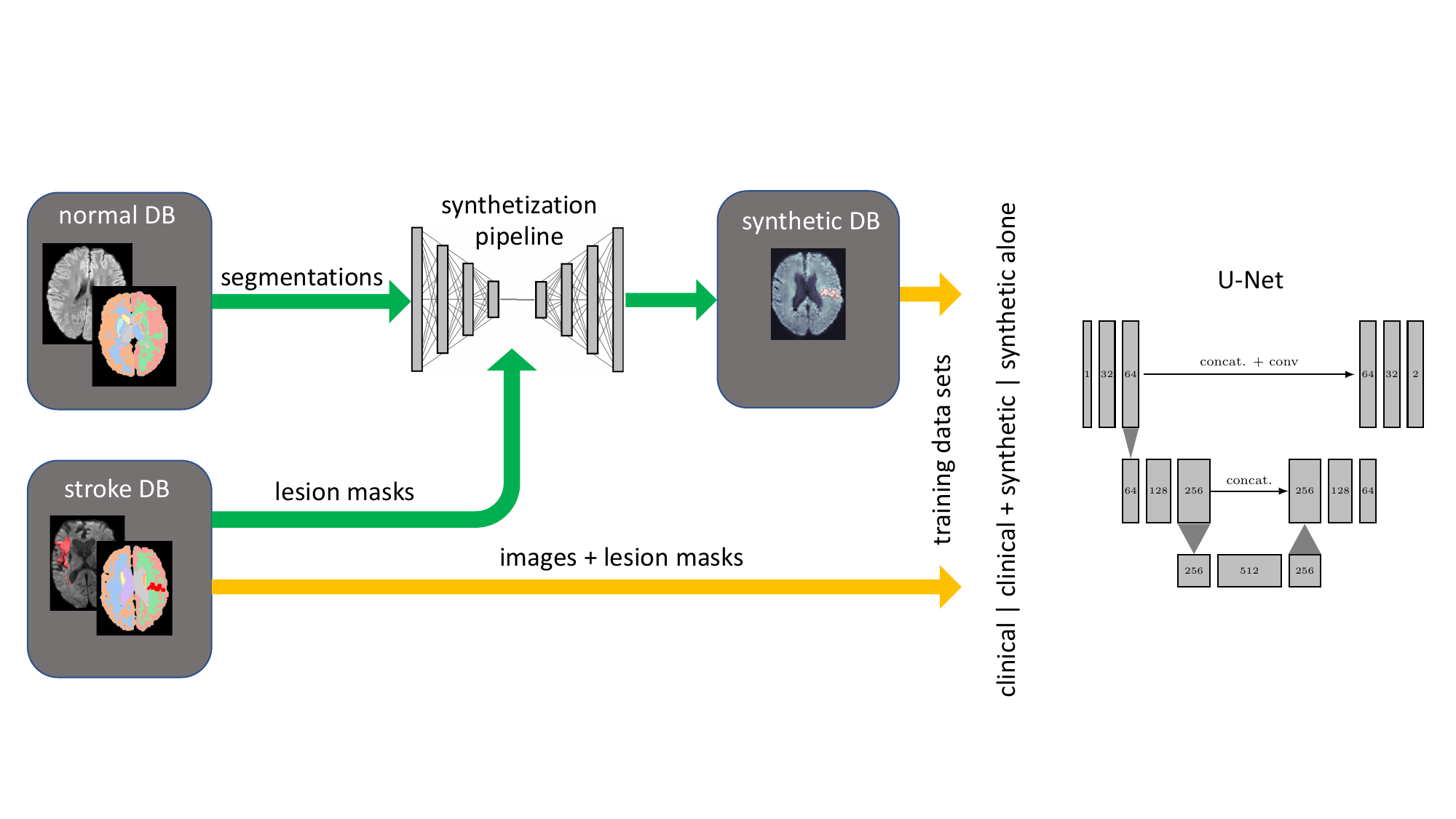}
\caption{Evaluation method. Once all ITMs are trained, each of them is used to generate a set of synthetic DWIs which are used as part of the training sets for the U-Net segmentation network. Ultimately, the U-Net is evaluated on a test set of clinical cases. The synthetization pipeline is shown in the graphical abstract.}
\label{fig:validation_procedure}
\end{figure*}
\subsubsection{SPADE} 
SPADE is the most recent of the ITMs considered in this work. While it has not been widely applied in the medical field, results on ambiental images are promising~\cite{SPADE}. Furthermore, the SPADE architecture, which is largely identical to a standard GAN, is very economical, reducing the number of parameters to train compared to other ITMs. In analogy to GANs~\cite{Goodfellow2014generative}, this model draws a random latent vector which is subsequently upsampled introducing elements of variation in the generative process. The segmentation mask is injected into SPADE normalization layers, which replace the usual batch- or instance-wise normalizations. This guides the model to learn more effectively and converge more rapidly to an optically appealing result~\cite{SPADE}.

\subsubsection*{Training}
We have implemented all ITMs as 2D CNNs in tensorflow \cite{TF} and trained  them on a single GPU (Nvidia Titan RTX 24GB) for 100 epochs on the combined clinical normal and stroke databases. All models used a batch size of 8, a learning rate of $2\cdot 10^{-4}$ and otherwise are set up as suggested in the references indicated in the previous section. All models are trained on equal footing, using the available segmented volumes to obtain a conditional generative model that differs from its competitors only in the model architecture, but not the training procedure.

Finally, we train a GAN, which serves as the lesion label generator, for 500 epochs on the 3D lesion masks that are part of the training set. The same lesion generator is used for all ITMs in the following lesion injection.

\subsection{Lesion injection and DWI synthesis}
In order to synthesize IS data to augment existing clinical data, we use the 2027 healthy brain DWIs' anatomical segmentation maps. 
In principle, one could also introduce synthetic segmentation maps, which entails complications of its own (mostly related to 3D generative modelling, see Sec.~\ref{sec:discussion}). Since this is not the main aspect of our analysis, we resort to the available segmentations of the healthy patient cohort. This also guarantees sufficient anatomical variation for the augmentation.
These healthy segmentation maps are subsequently modified by injecting a fake lesion, which in turn is generated by a 3D Wasserstein GAN~\cite{Arjovsky2017WassersteinG} that is trained on the lesion masks of the stroke DB and whose softmax outputs are transformed into hard segmentations by thresholding at a value of 0.5. The injection is achieved by adapting the label in the parenchyma of the DWI's segmentation map; however, demanding a minimum lesion volume of 20 voxels. If this requirement is not met, a new lesion mask is generated until a lesion of at least 20 voxels is generated.\footnote{We have also investigated smaller thresholds, but found no improvement, as this tends to increase the false-positive rate of the segmentation networks.}
This approach ensures that the anatomical features of the generated brain volumes are realistic and we do not generate a large amount of small, scattered lesions. The Wasserstein loss has proven to be most stable in training. Moreover, it is assumed that the lesion generation by a GAN will generalize previous attempts to data augmentation by synthetic data, e.g.~Ref.~\cite{FederauStroke}. The underlying assumption is that generative models yield meaningful interpolations in latent space and at the same time generalizations of simple (geometric) transformations of the available clinical lesion masks.

The generated segmentation map is subsequently decomposed into 2D slices and fed into the ITM in order to generate a fake DWI. In this manner we obtain a database of 2027 fake DWIs per ITM.
Fig.~\ref{fig:implantation} shows an example of this lesion implantation procedure, which is also sketched in the graphical abstract. The left column shows the healthy input DWI volume, whose corresponding semantic segmentation map is shown next to it. The third column shows the segmentation map after implanting a lesion label according to the output of the GAN. 
Subsequently, we have generated realistic DWIs from axial slices of these 3D segmentation maps using three ITMs, namely Pix2Pix, SPADE and cycleGAN, as shown in the right-most column for one axial slice. 

\subsection{Evaluation of lesion segmentation}\label{sec:eval}
In order to assess the performance of the various ITMs quantitatively, we train a 3D U-Net~\cite{RonnebergerUNet} segmentation network on the clinical, clinical and synthetic, and synthetic-alone databases (for each ITM discussed above) and subsequently evaluate the resulting segmentation masks on a test set comprising 74 stroke patient and 85 healthy patient DWIs, which has been separated from the training data. See Fig.~\ref{fig:validation_procedure} for a visual representation of this procedure.

The U-Net is a well-established standard in biomedical image segmentation and has been used for various tasks, including stroke~\cite{FederauStroke,ZhangStroke}, and still yields results that match the state-of-the-art~\cite{nnUnet}. Training is stopped if the validation loss has not increased for 100 epochs. To this end, 15\% of the training data are used for validation.
To compare the models' performance quantitatively, we consider a range of 100 epochs, chosen at the end of the training history, to evaluate each model on the test set. 

The U-Net architecture is outlined in detail in the Appendix. 
The network's output is a feature map of equal dimension as the input; however, comprising as many image channels as segmentation labels (in our case two: background and lesion), from which a hard segmentation can be obtained by applying an argmax over the channel axis. 

The models are trained on a combination of cross-entropy loss and DSC, which are computed using the soft-max output of the U-Net $P$ and the ground truth $G$ as
\begin{gather}
    \mathcal{L}_\text{U-Net}(P, G) = - DSC(P, G) + \sum_i G_i\log  P_i \\
    \mathrm{DSC}(P, G) \equiv \frac{2\,| P \cap G|}{|P| + |G|} = \frac{2\,\sum_i P_i G_i}{\sum_i P_i + \sum_j G_j},\label{eq:dice}
\end{gather}
where $|\cdot|$ denotes the cardinality of a set and the second equality in \eqref{eq:dice}
holds for a pixel-/voxel-based binary segmentation map, and the sums run over all entries. For better comparability, we also report the DSC between the segmentation by two human
readers (each with 2 years of experience). 

Additionally, we evaluate the quality of the segmentations by a number of additional metrics. First, we evaluate the relative volume difference, given that the lesion volume is an important quantity to triage patients for treatment. Further, we evaluate the Hausdorff distance (HD) and the average symmetric surface distance (ASSD), two quantities that are often reported in order to quantify the quality of the predicted lesion shapes compared to the ground truth. The surface distance $d_A(b)$ is defined as the minimal Euclidean distance of point $b \in B$ from surface $A$, i.e.~the minimal distance between $b$ and any point $a\in A$. From this, the ASSD is constructed as the symmetric sum,
\begin{align}
\mathrm{ASSD}(A, B) = \frac{1}{|A| + |B|} \left( \sum_{a \in A} d_B(a) + \sum_{b \in B} d_A(b) \right),
\end{align}
while the (symmetrized) HD is the largest symmetric surface distance,
\begin{align}
\mathrm{HD}(A, B) =\frac{1}{2} \left( \max_{a \in A} d_B(a) + \max_{b \in B} d_A(b) \right).
\end{align}
Finally, we evaluate recall and precision of the models. All calculations are relying on the python package \texttt{MedPy}~\cite{medpy}.

The U-Net is trained on clinical, synthetic, and combined data sets for 500 epochs (clinical data alone), or 300 epochs (all others) using a batch size of 5 and a learning rate of $10^{-4}$. To ensure the best benchmark possible, we use standard affine data augmentation including flips, rotations, sheer and translation for the clinical data~\cite{DataAug}. In some cases, we further investigate if the performance can be improved by fine-tuning the model trained on synthetic cases using only a handful of clinical cases.

In order to assess the impact of the training set sizes, we have conducted a number of follow-up experiments, where we evaluate the metrics discussed above and calculate mean and standard deviation over the test set for models trained on fractions of the clinical data and the synthetic data. We consider two cases: One, where we used the normal DB's segmentation masks to synthesize fake data, and one where we used the 428 DWI-negative stroke DB cases instead. Finally, we use these 428 cases' segmentations four times to inject lesions with different random seeds to generate a synthetic database of size comparable to that of the normal DB.
Moreover, we analyze the impact of different conventional data augmentation techniques, and using no augmentation at all.

\begin{figure}[t]
    \centering
    \begin{minipage}{0.15\textwidth}\centering
    \includegraphics[width=\textwidth, trim={30cm 51cm 5cm 43.5cm}, clip]{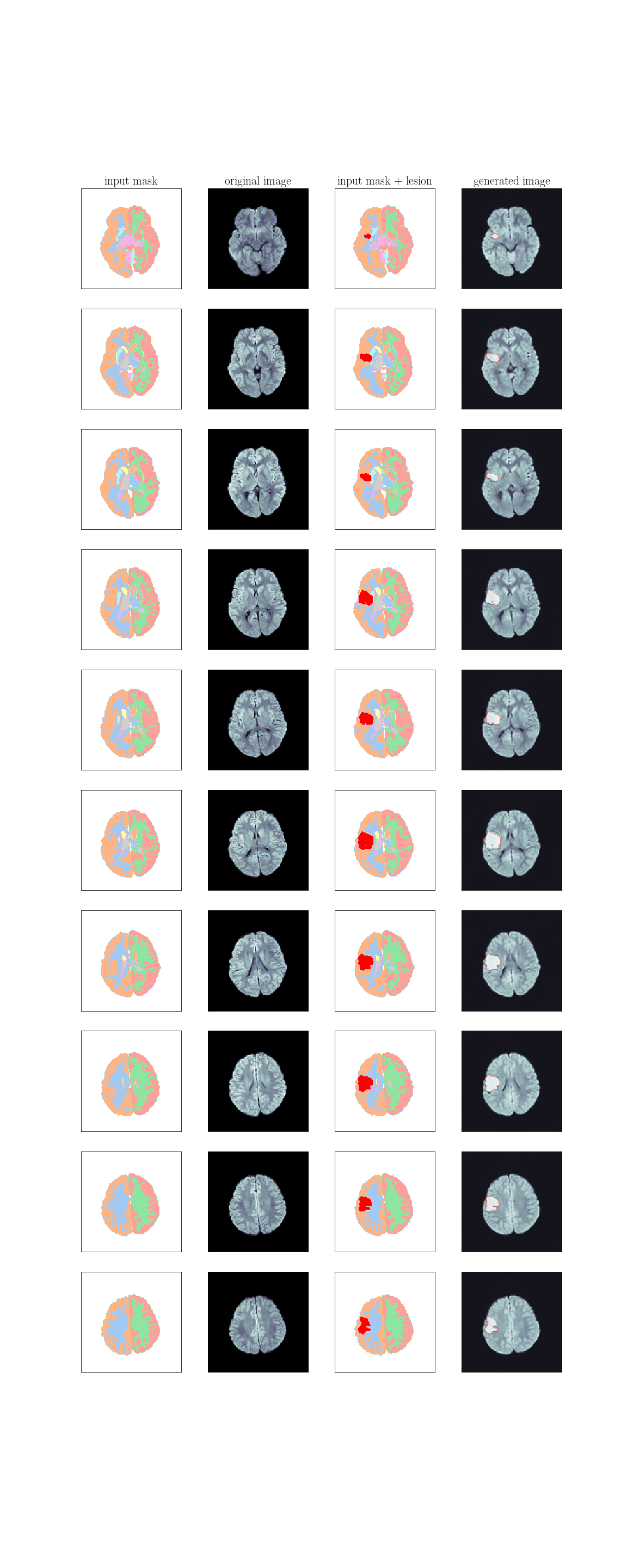}
    \\
    \mbox{\small(a) Pix2Pix ($\lambda=10$)}
    \end{minipage}
    \hspace{1mm}
    \begin{minipage}{0.15\textwidth}\centering
    \includegraphics[width=\textwidth, trim={30cm 51cm 5cm 43.5cm}, clip]{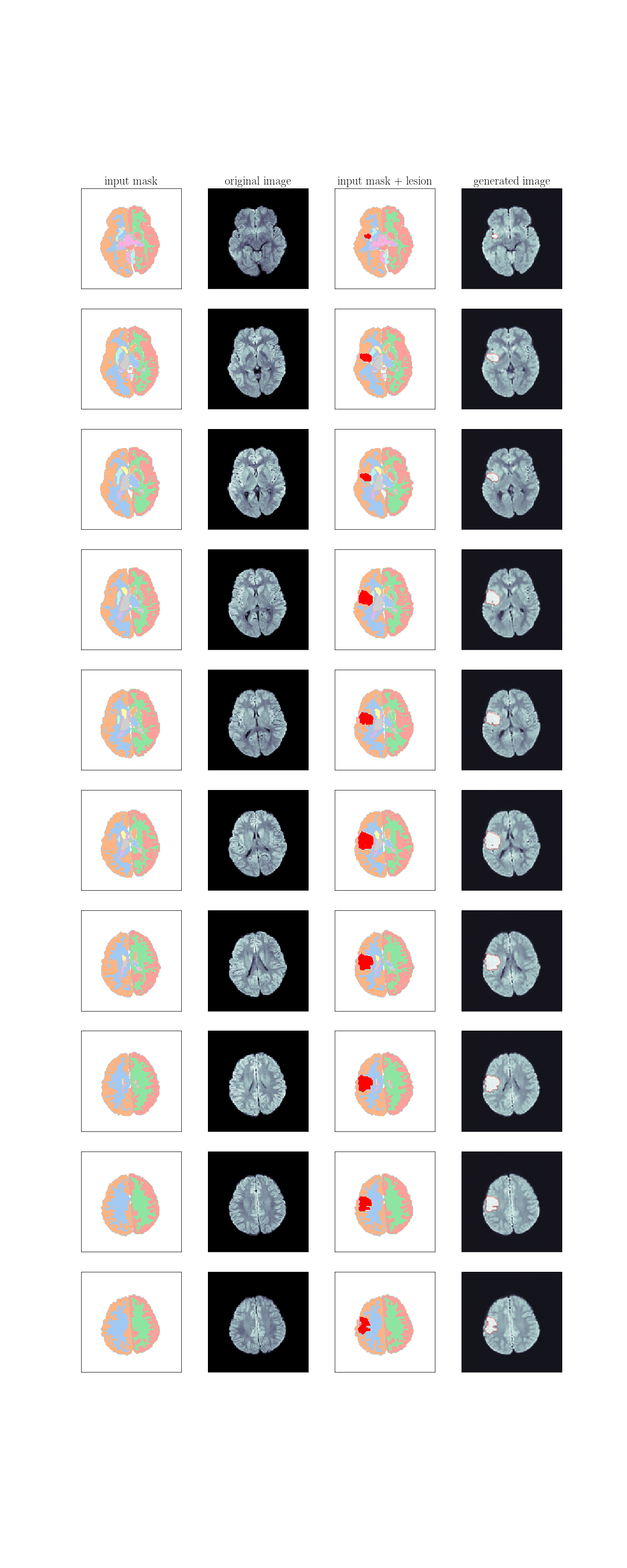}
    \\
    \mbox{\small(b) Pix2Pix ($\lambda=100$)}
    \end{minipage}
    \\
    \vspace{2mm}
    \begin{minipage}{0.15\textwidth}\centering
    \includegraphics[width=\textwidth, trim={30cm 51cm 5cm 43.5cm}, clip]{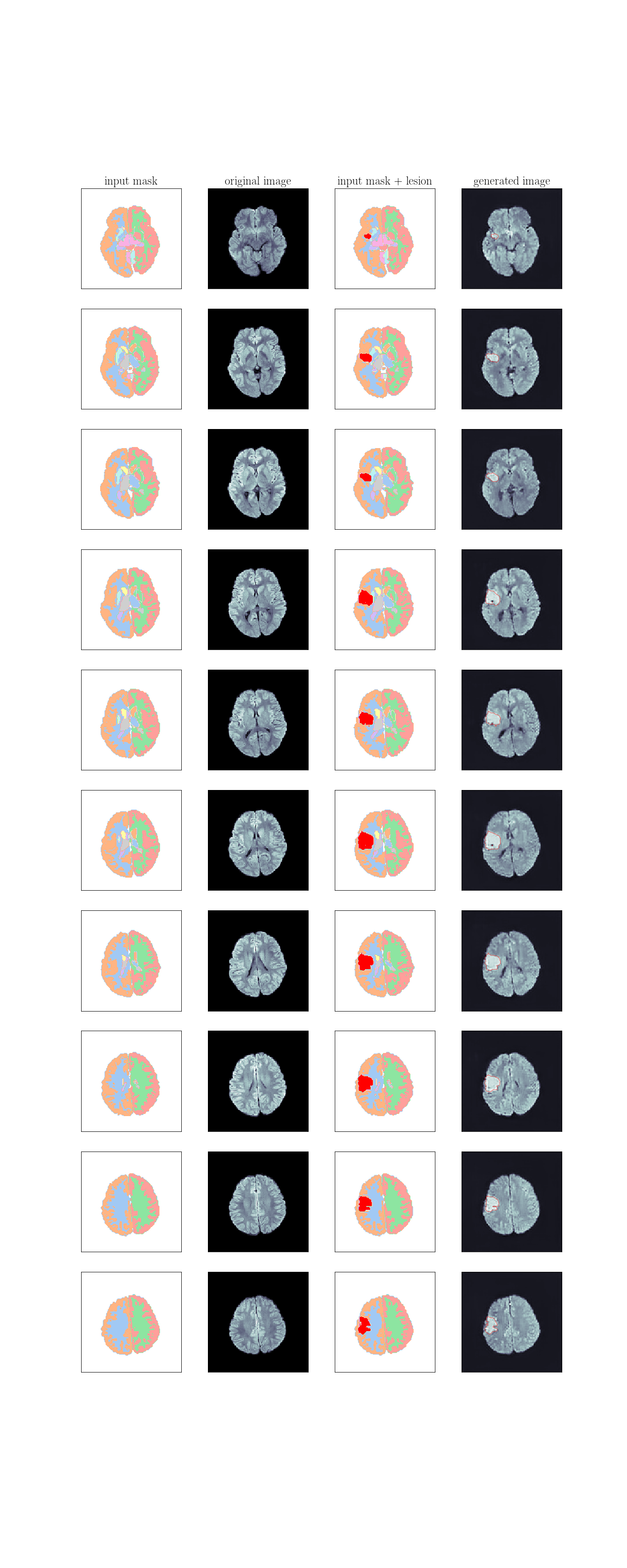}
    \\
    \mbox{\small(c) SPADE}
    \end{minipage}
    \hspace{1mm}
    \begin{minipage}{0.15\textwidth}\centering
    \includegraphics[width=\textwidth, trim={30cm 51cm 5cm 43.5cm}, clip]{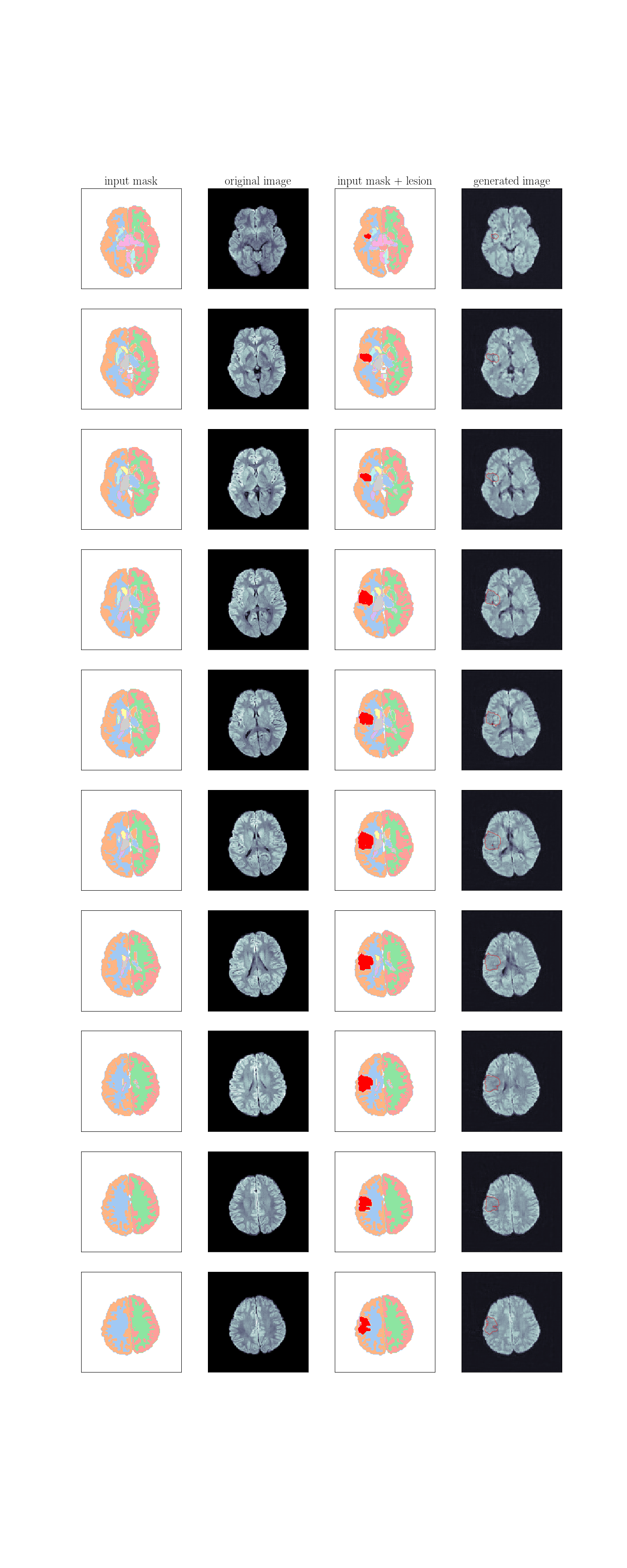}
    \\
    \mbox{\small(d) cycleGAN}
    \end{minipage}
    \caption{Generated stroke DWIs from the segmentation mask in Fig.~\ref{fig:implantation}.}
    \label{fig:generation}
\end{figure}

\begin{figure}[b]
    \centering
    \includegraphics[width=.85\linewidth]{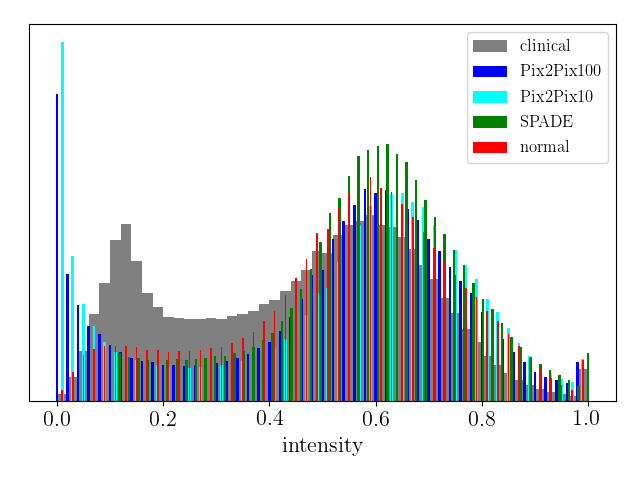}
    \caption{Intensity distribution of the synthetic and clinical stroke images.}
    \label{fig:intensity_histograms}
\end{figure}

\section{Results}\label{sec:results}
\subsection{Synthetic data generation and qualitative evaluation}\label{sec:data_synthesis}
Fig.~\ref{fig:generation} shows some examples of generated images. The synthetic DWIs generated by the ITMs display anatomical structures that are coherent in location, shape, and size across the volume, including the lesions. 
The first two images from the left have been generated using Pix2Pix. Panel~(a) was generated using a reconstruction loss weight $\lambda=10$, while Panel~(b) was created using $\lambda=100$. The results are of good quality, subjectively realistic and, as anticipated, come accompanied by a high-fidelity lesion label. Only closer inspection of a number of volumes reveals that the larger reconstruction loss weight produces slightly sharper images with more details both inside the lesion and in the brain tissue. We have therefore tested both models quantitatively in the following section.

The lesion intensity produced by SPADE [Fig.~\ref{fig:generation}(c)] is noticeably lower than in the images generated by Pix2Pix, but also the contrast outside the lesion is much lower: in the posterior part of the synthetic brain, barely any gray/white matter contrast is present. While this might seem a disadvantage at first sight, it could actually be beneficial for training a U-Net to reliably detect lesions with lower signal increase.

Next, notice that cycleGAN has failed to recognize and synthesize any lesion hyper-intensity, while producing very high inter-tissue contrast [Fig.~\ref{fig:generation}(d)]. We have experimented with fine-tuning the model on IS data, training on IS data alone, and modifying the network architecture; however, this result seems to be unavoidable. We speculate that this is closely related to the cycle consistency requirement of cycleGAN, i.e.~that the two generators map an image back to itself as close as possible, and large class imbalance between lesion and non-lesion voxels. Similarly, Ref.~\cite{cycleGAN} identified `failed cases' which indicate that cycleGAN generalizes poorly.
Due to this failure, we have excluded cycleGAN from the quantitative analysis, and leave a dedicated study using cycleGAN for data augmentation for future work. We stress that this does not preclude any application of cycleGAN to this task; however, it will require an approach that differs from the one chosen here, and thus the comparability of ITMs is not given.

To conclude this section, we present in Fig.~\ref{fig:intensity_histograms} the intensity histograms of the synthetic and real data sets after pre-processing and normalized to the range $[0,1]$. Notice how the normal and clinical (i.e.~stroke) DBs significantly differ in the region of intensities below $0.5$. Moreover, the synthetic image intensities closely follow the normal volume intensities. This does not come as a surprise as we have used the normal segmentation masks to synthesize stroke images. One is led to the conclusion that some systematic difference, either of anatomical nature or rooted in the image acquisition, is responsible for this discrepancy. We will now turn to the assessment of the quality of image generation and return to this question in Sec.~\ref{sec:discussion}.

\begin{figure}[t]
    \centering
    \includegraphics[width=.85\linewidth]{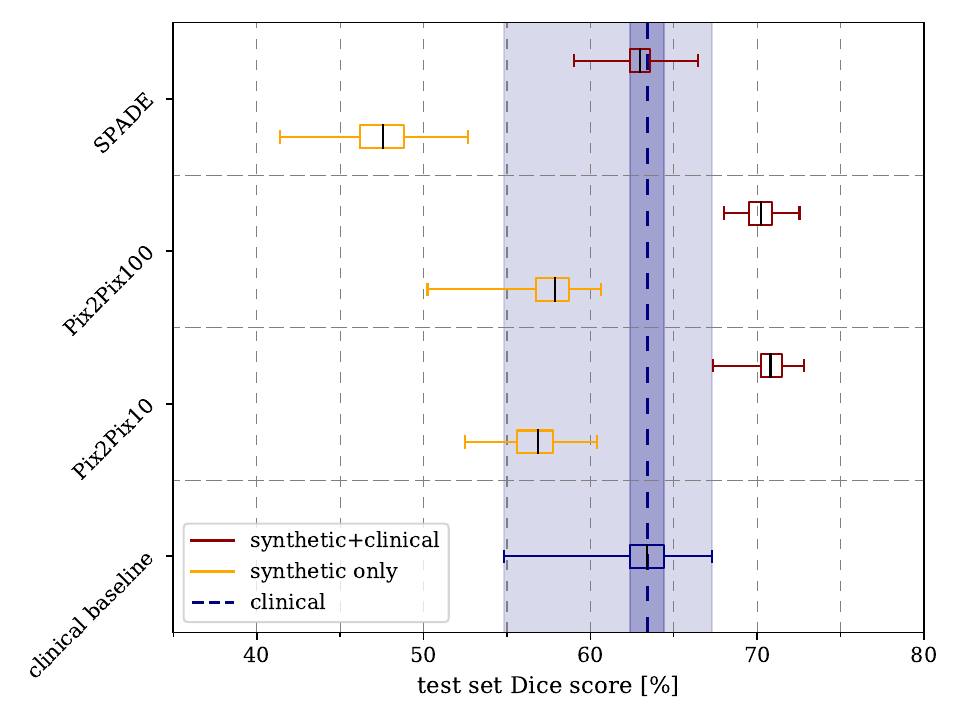}
    \caption{Box plot of the various ITMs evaluated on the clinical test set. Boxes indicate standard deviation and whiskers min- and max-values.}
    \label{fig:model_comparison}
\end{figure}
\subsection{Model selection and test set evaluation}\label{sec:evaluation}
The results are shown in Fig.~\ref{fig:model_comparison} and summarized in Tab.~\ref{tab:results}. In our set-up, the U-Net trained on the clinical data reaches a mean test set DSC of $(63.2 \pm 1.9) \%$ and a maximal DSC of 67.3\%. The U-Nets trained on the combined synthetic data from Pix2Pix and clinical data outperform the clinical baseline U-Net and we find mean DSCs of $(70.3 \pm 1.1)\%$ (Pix2Pix; $\lambda = 100$) and $(70.8 \pm 1.0) \%$ (Pix2Pix; $\lambda=10$), while SPADE performs slightly worse ($62.9 \pm 1.2$). The maximum DSC achieved is $72.8\%$ for Pix2Pix with $\lambda=10$. The other models achieve maximal DSC of $66.5\%$ (SPADE) and $72.5\%$ (Pix2Pix; $\lambda=100$). This is a key result of our study highlighting that synthetic data does not simply reproduce available information in the training data, but instead interpolates and effectively generalizes the training data.

In contrast to this, the U-Net performs much worse when trained on synthetic data alone, and with much greater variance. We find the test set DSCs $(57.1 \pm 2.7) \%$ (Pix2Pix; $\lambda=100$), $(55.7 \pm 3.3)\%$ (Pix2Pix; $\lambda = 10$), and $(47.3 \pm 2.2)\%$ (SPADE). Interestingly, and different from the previous result, the larger reconstruction loss yields better results for Pix2Pix in agreement with our qualitative observation. 
To put these findings into perspective we report the human inter-reader DSCs of $76.6 \pm 13.9 \%$ for the training and $76.9 \pm 13.5 \%$ for the test set.

\begin{figure}[b]
    \centering
    \includegraphics[width=.85\linewidth]{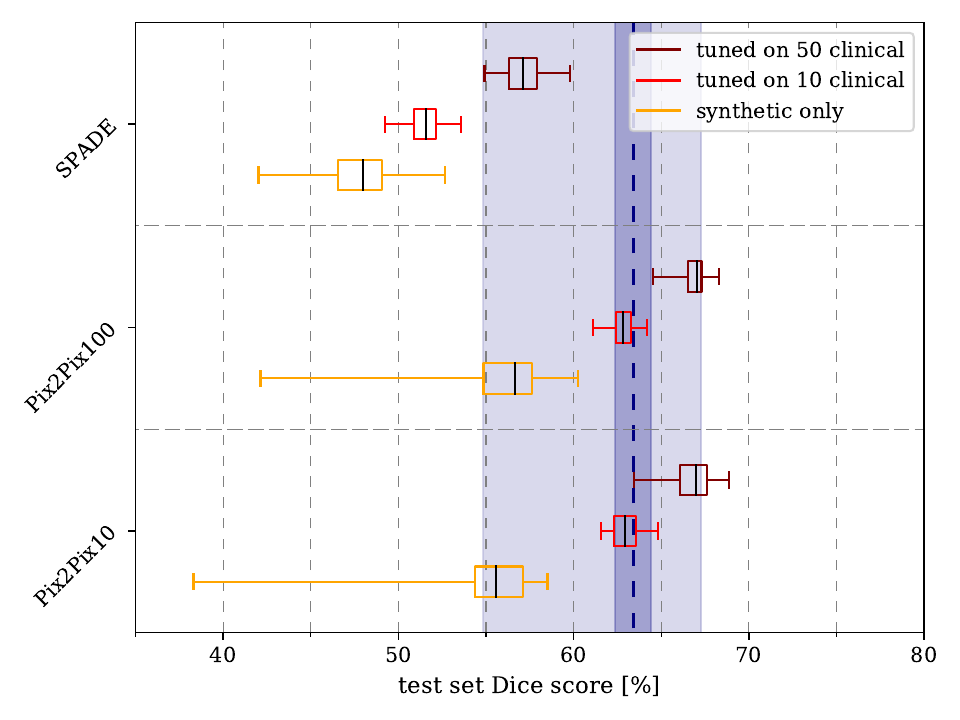}
    \caption{Box plots for fine-tuned models. The performance of models trained on synthetic data alone can be improved to match the clinical case, or even outperform it in some cases.}
    \label{fig:model_tuning}
\end{figure}

\subsection{Model fine-tuning} \label{sec:finetuning}

These results indicate that synthetic data alone is
not apt to replace the clinical data set. Whether this can be attributed to the quality of the generated images or differences in anatomical features between the databases used for image synthesis and evaluation, will be investigated in the upcoming section. First, we address the question whether the models trained on synthetic data alone can be fine-tuned to yield better results.

To this end, we fine-tune a model previously trained on synthetic data alone on either 10 or 50 randomly chosen clinical training samples for another 150 epochs. In Fig.~\ref{fig:model_tuning} we duplicate the previous result, where the model was trained on synthetic data alone (orange). Subsequently, we show the results after training on 10 (red) and 50 clinical cases (dark red), respectively.  
Notice the difference between the two distinct architectures, Pix2Pix and SPADE. Increasing the tuning set size from 10 to 50 yields only a minor improvement for SPADE: from $(51.0 \pm 1.2)\%$ for 10 samples to $(56.4 \pm 1.5)\%$ for 50 samples. After tuning on 50 samples, SPADE's performance matches that of the lower spectrum of the clinical baseline model. The situation is quite distinct for Pix2Pix, which for 10 tuning samples matches the performance of the clinical baseline. Tuning on 50 samples yields enough information for Pix2Pix to outperform the clinical model with a mean DSC of $(66.4 \pm 1.1)\%$ (max.~$68.9\%$; $\lambda=10$) and $(66.8 \pm 0.7)\%$ (max.~$68.3\%$; $\lambda=100$). In all cases the variance of DSCs is much lower compared to the clinical baseline.Had we trained the U-Net on only 10 [50] cases, the outcome would be much worse: we find a mean DSC of $(2.9 \pm 1.6)\%$ [$(12.5 \pm 7.0)\%$]. 

Therefore, we may conclude that models trained on synthetic data alone do not match the performance of those trained on clinical data sets. However, these models can be fine-tuned with only few clinical training cases. This ultimately yields results that are comparable to the clinical baseline model trained on all 365 clinical IS cases; however, with much less variance. Moreover, in a situation where only a limited number of clinical cases is available, this approach yields a significant improvement as we highlight in the upcoming section. 

In Fig.~\ref{fig:examples} we show some examples for segmentations on two clinical test cases, one with a big lesion (top row) and one with small, disconnected lesions (bottom row). The U-Nets have been trained on clinical data (left-most panels), or on the combined data sets as indicated (other panels). Alongside the labeled areas, the ground truth (red) and the corresponding DSC (for the whole volume) are indicated. 

\subsection{Further experiments}

\begin{figure}[t]
    \centering
    \includegraphics[width=\linewidth, trim= 0 5mm 0 5mm, clip]{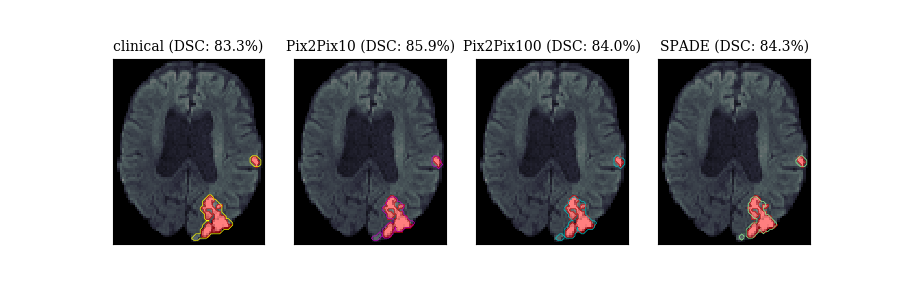}\\
    \includegraphics[width=\linewidth, trim= 0 5mm 0 5mm, clip]{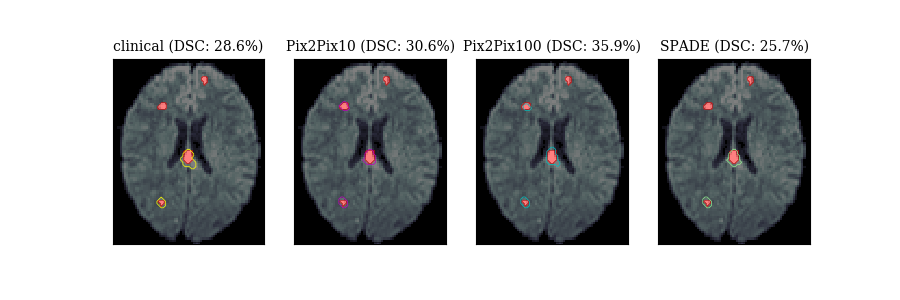}
    \caption{Two segmentation examples: A large lesion (\textit{top}), which is segmented well by all models. And smaller, disconnected lesions (\textit{bottom}). The latter case shows greater DSC variance among the models.}
    \label{fig:examples}
\end{figure}

\begin{figure*}[t]
    \centering
    \includegraphics[width=0.75\textwidth]{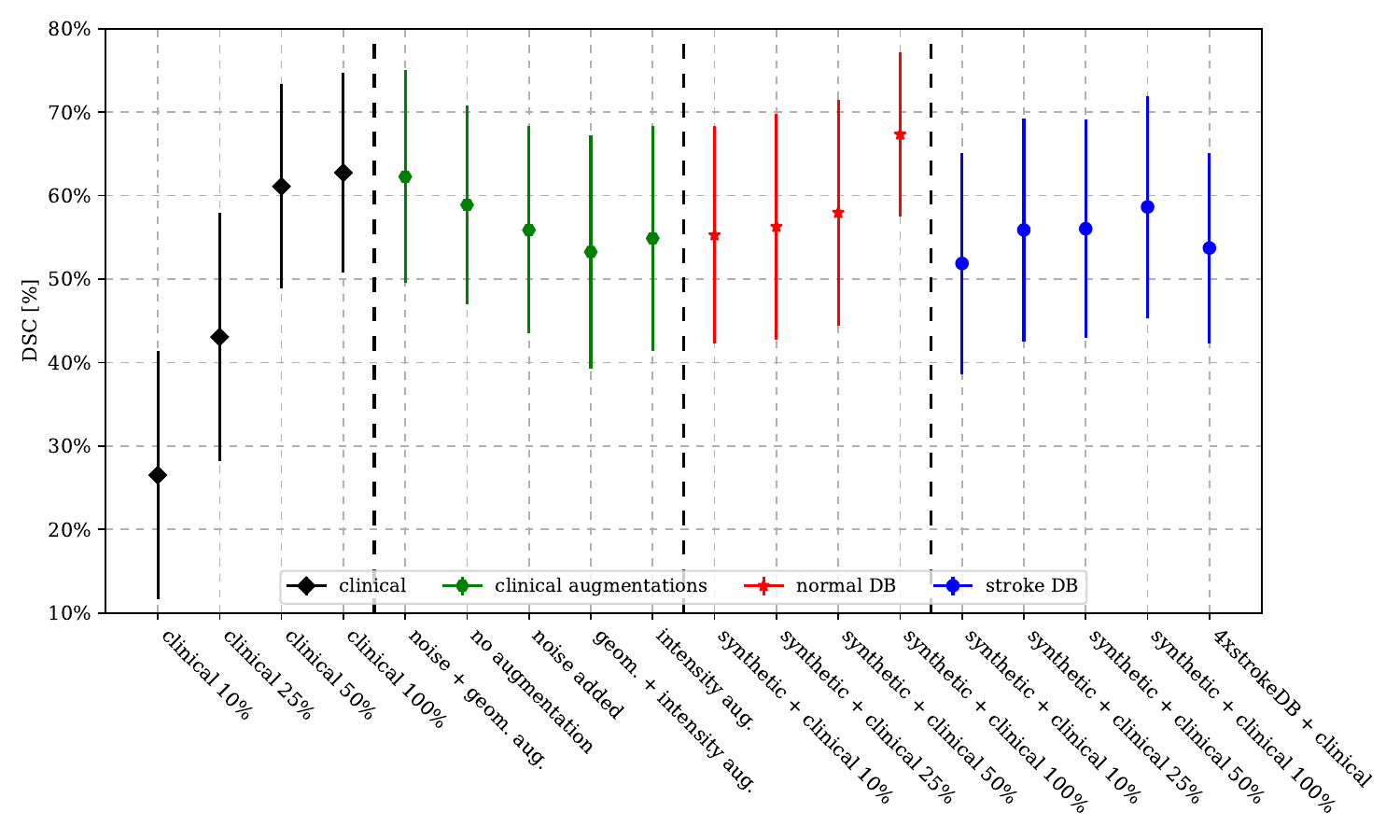}
    \caption{Results of the additional experiments, which correspond to the first column of Tab.~\ref{tab:frac_data}.}
    \label{fig:experiments}
\end{figure*}

Let us now focus on the model that obtained the highest DSC in Sec.~\ref{sec:evaluation}, i.e.~Pix2Pix ($\lambda=10$) and consider it in more detail. Tab.~\ref{tab:frac_data} and Fig.~\ref{fig:experiments} show the results of the four different types of experiments we have conducted.
The first for rows contain the results of training the U-Net on a fraction of $10/25/50\%$ of the clinical data and evaluated on the full test set.
Notice that in all cases, the error is rather large and decreases only slightly when more clinical data is included, indicating that the data set is rather diverse. The DSC increases quickly as the training set size grows; however, only a small improvement is found when going from 50\% to the full data set. This behavior is contrasted by the relative volume loss, which is in fact largest for the full training set. Closer inspection reveals that cases with large volume loss have small lesion volumes (see also Fig.~\ref{fig:examples}), and thus the volume loss does not seem to be a suitable metric when lesions of varying sizes are studied. All other metrics confirm that using the full training set yields the best performance.

In the second class of experiments, we utilized the full training set and used different data augmentation techniques, including adding random noise and randomly remapping the image intensities as suggested in Ref.~\cite{intensityAug}.\footnote{Recall that, unless indicated otherwise, we always use geometrical data augmentation.} However, none of the metrics indicates that any other data augmentation technique performs better than geometrical transformation alone.

\begin{table}[b]
\centering
\begin{tabular}{lcccc}
    data set    &   baseline    &   Pix2Pix10    &   Pix2Pix100    &   SPADE\\
        \toprule
    clinical    &   $\mathbf{63.2 \pm 1.9}$   &   $-$   &   $-$   &   $-$ \\
    + all synthetic   &  $-$   &   $\mathbf{70.8 \pm 1.0}$ &   $70.3 \pm 1.1$  &    $62.9 \pm 1.2$\\
    synthetic   &   $-$ &   $55.7 \pm 3.3$ &   $\mathbf{57.1 \pm 2.7}$  &   $47.3 \pm 2.2$\\
    + 10 clinical   &   $-$ &   $\mathbf{62.9 \pm 0.8}$  &   $62.3 \pm 0.9$  &   $51.0 \pm 1.2$\\
    + 50 clinical   &   $-$ &   $66.4 \pm 1.1$  &   $\mathbf{66.8 \pm 0.7}$  &   $56.4 \pm 1.5$ \\
    \bottomrule
\end{tabular}
\caption{Summary of the resulting DSCs (in percent) of the different models trained with various amounts of synthetic data. The last two rows correspond to the fine-tuning procedure described in Sec.~\ref{sec:finetuning}.}
\label{tab:results}
\end{table}

\begin{table*}[t]
\begin{tabular}{lcccccc}
\toprule
model &              DSC [\%] &  rel. volume loss [\%] &               HD [mm] &             ASSD [mm] &        precision [\%] &           recall [\%] \\
\midrule
10\% clinical                            &  $26.5 \pm 29.7$ &    $58.2 \pm 0.3$ &  $42.7 \pm 32.7$ &  $14.0 \pm 20.9$ &  $52.7 \pm 42.2$ &  $21.2 \pm 26.5$ \\
25\% clinical                            &  $43.1 \pm 29.7$ &    $42.9 \pm 0.3$ &  $40.5 \pm 30.2$ &   $8.6 \pm 16.0$ &  $68.1 \pm 32.7$ &  $38.3 \pm 31.8$ \\
50\% clinical                            &  $61.1 \pm 24.5$ &    $44.6 \pm 2.4$ &  $47.6 \pm 29.3$ &   $7.8 \pm 16.6$ &  $71.3 \pm 22.7$ &  $57.8 \pm 27.0$ \\
100\% clinical                           & $62.7 \pm 23.9$ &    $78.7 \pm 3.5$ &  $33.2 \pm 25.5$ &   $4.7 \pm 10.7$ &  $75.6 \pm 22.5$ &  $59.6 \pm 27.2$ \\
\midrule
geometrical augment + noise                      &  $62.3 \pm 25.5$ &    $86.5 \pm 3.5$ &  $35.3 \pm 28.4$ &   $6.5 \pm 14.8$ &  $74.7 \pm 24.8$ &  $59.4 \pm 28.4$ \\
no augmentation                                 &  $58.9 \pm 23.8$ &    $51.1 \pm 1.6$ &  $38.3 \pm 26.0$ &   $6.3 \pm 13.9$ &  $73.7 \pm 22.4$ &  $54.7 \pm 27.8$ \\
noise                                   &  $55.9 \pm 24.8$ &  $166.4 \pm 11.4$ &  $43.4 \pm 28.5$ &   $6.4 \pm 11.7$ &  $72.3 \pm 24.9$ &  $51.0 \pm 27.1$ \\
geometrical + intensity augmentation                  &  $53.3 \pm 28.0$ &    $33.1 \pm 0.3$ &  $34.9 \pm 28.4$ &   $6.5 \pm 12.7$ &  $80.1 \pm 25.4$ &  $45.1 \pm 28.5$ \\
intensity augmentation                          &  $54.9 \pm 26.9$ &  $387.2 \pm 26.7$ &  $42.2 \pm 25.7$ &   $8.0 \pm 16.7$ &  $71.4 \pm 26.9$ &  $51.0 \pm 29.1$ \\
\midrule
10\% clinical + Pix2Pix (normal DB)      &  $55.3 \pm 26.0$ &    $29.9 \pm 0.3$ &  $38.6 \pm 29.1$ &   $7.0 \pm 13.1$ &  $79.2 \pm 24.7$ &  $46.6 \pm 26.4$ \\
25\% clinical + Pix2Pix (normal DB)      &  $56.3 \pm 27.1$ &    $27.9 \pm 0.3$ &  $39.4 \pm 28.0$ &   $7.5 \pm 16.4$ &  $79.8 \pm 25.0$ &  $48.5 \pm 28.2$ \\
50\% clinical + Pix2Pix (normal DB)      &  $58.0 \pm 27.1$ &    $28.3 \pm 0.3$ &  $36.6 \pm 29.5$ &   $6.3 \pm 14.2$ &  $79.1 \pm 24.4$ &  $51.3 \pm 28.9$ \\
100\% clinical + Pix2Pix (normal DB)     &  $67.3 \pm 19.7$ &    $15.5 \pm 0.2$ &  $32.2 \pm 26.4$ &   $4.1 \pm 10.6$ &  $81.8 \pm 15.6$ &  $61.8 \pm 23.9$ \\
\midrule
10\% clinical + Pix2Pix (stroke DB)      &  $51.9 \pm 26.5$ &    $27.4 \pm 0.3$ &  $47.6 \pm 28.8$ &   $8.1 \pm 11.7$ &  $72.8 \pm 24.9$ &  $44.6 \pm 27.3$ \\
25\% clinical + Pix2Pix (stroke DB)      &  $55.9 \pm 26.7$ &    $27.5 \pm 0.3$ &  $42.5 \pm 27.9$ &    $5.2 \pm 6.2$ &  $76.9 \pm 20.7$ &  $50.4 \pm 29.2$ \\
50\% clinical + Pix2Pix (stroke DB)      &  $56.0 \pm 26.2$ &    $35.3 \pm 0.6$ &  $38.0 \pm 26.0$ &    $5.6 \pm 8.4$ &  $77.1 \pm 20.8$ &  $50.9 \pm 29.9$ \\
100\% clinical + Pix2Pix (stroke DB)     &  $58.6 \pm 26.6$ &    $41.8 \pm 1.0$ &  $37.3 \pm 29.3$ &   $7.0 \pm 14.4$ &  $77.0 \pm 25.1$ &  $54.1 \pm 29.7$ \\
100\% clinical + 4$\times$Pix2Pix (stroke DB) &  $53.7 \pm 22.8$ &   $137.7 \pm 8.9$ &  $54.6 \pm 26.5$ &   $8.7 \pm 10.3$ &  $68.3 \pm 26.3$ &  $48.6 \pm 23.4$ \\
\bottomrule
\end{tabular}
\caption{Experiments with varying training set sizes and compositions. Results indicate mean and standard deviation over the test set The first section compares models with fractional clinical data as indicated. The second section compares different data augmentations. In the third and forth column, the results using synthetic data from Pix2Pix are shown ($\lambda=10$) synthesized from the normal DB (third column) and the stroke DB (fourth column). In each section, results are ordered by increasing DSC and we include the additional metrics discussed in Sec.~\ref{sec:methods}.}
\label{tab:frac_data}
\end{table*}

In the former case, significant improvements can be obtained for the small fractional training data sets, e.g.~with less than 40 DWI-positive clinical cases (10\%), a DSC above 50\% can be obtained. Other metrics indicate the same behavior. Interestingly, the relative volume loss is monotonically decreasing with increasing number of clinical cases, indicating that the model's predictions are more stable for smaller lesions. 

One might have suspected that using segmentation masks from the stroke DB (which on average have larger ventricles due to the higher age of the individuals), would further improve the performance; however this is not the case. Even when recycling the segmentation masks four times (which yields a data set of size equal to the normal DB) and augmenting with different lesion labels, we do not find any improvement. Thus, we may conclude that the increased variability in the normal DB is beneficial to the performance of the U-Net. Moreover, we conclude that the deficiencies of the models trained on synthetic data alone are not rooted in the out-of-distribution nature of the segmentation masks we used, but rather a limitation of the ITMs themselves, or the training set size, respectively.

\begin{figure}
    \centering
    \includegraphics[width=0.8\linewidth]{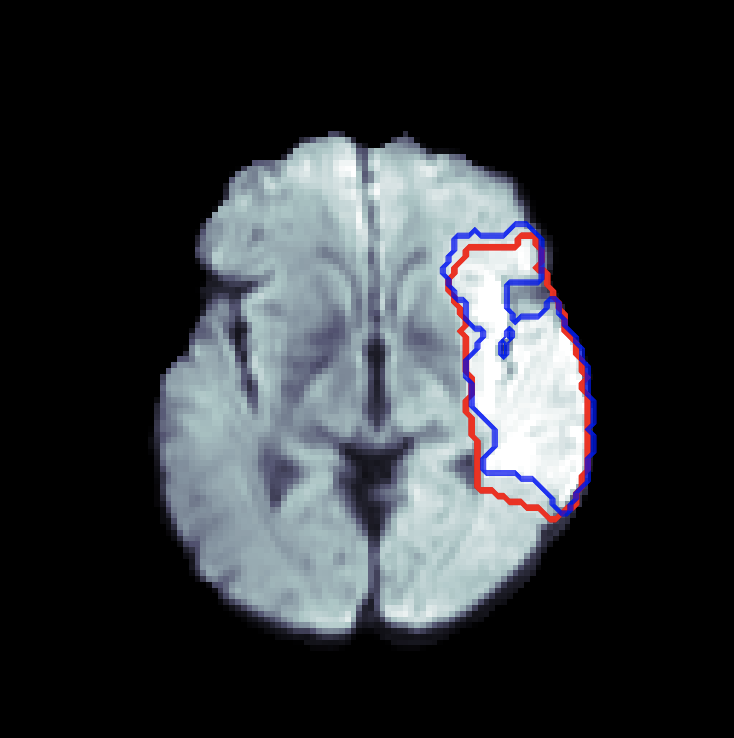}
    \caption{Segmentation result on a sample from the ISLES 2015 DWI training data. Red: ground truth, blue: model prediction. Notice the coarse ground truth where cerebrospinal fluid is labeled as belonging to a lesion, whereas the model prediction correctly follows the brain tissue boundaries.}
    \label{fig:ISLES}
\end{figure}

Finally, we have evaluated the trained models on the ISELS 2015 DWI images~\cite{ISLES}. We found that the models under consideration unanimously yield DSCs around 50\% with little variation among models. However, the variation across this test set is rather large with a standard deviation of around 30\%. Closer inspection of some of the cases reveals that the ISLES data is much more coarsely segmented, such that the DSCs are actually limited by the segmentation quality of this test set rather than the models' capacities. Qualitatively, the results are of high quality and only a few cases, with infraction in the brain stem are missed by our models. See Fig.~\ref{fig:ISLES} for an example, where the model provides a very detailed outline of the segmentation. The ground truth segmentation, on the other hand is very coarse, including even CSF and therefore structures that are not part of the parenchyma. This strongly supports our hypothesis that the quality of the data inhibits the scores of the models rather than the model performance itself.

\subsection{Discussion}\label{sec:discussion}

The results illustrate that DL-based data augmentation can be used to optimize model performance. Depending on the number of available clinical samples, the improvement can be substantial, as summarized in Tabs.~\ref{tab:results} and~\ref{tab:frac_data}. 

Training on the clinical set augmented by synthetic data, Pix2Pix with a reconstruction loss weight of $\lambda = 10$ is the choice that yields the best results. Training instead on synthetic data alone, we find that increasing the reconstruction loss weight to $\lambda = 100$ results in the highest DSC. With 10 training samples to fine-tune this model, it yields results comparable to the clinical data set, while it produces better segmentations with only 50 fine-tuning samples. This result matches the qualitative finding that a larger reconstruction loss yields sharper images than the recommended  $\lambda=10$; however, the differences are rather marginal.
A more economical model with less adjustable parameters is the SPADE framework, which is attractive for two reasons: First, less trainable weights signify less time for the model to converge and require less training data. And secondly, since SPADE samples from a latent space, it introduces elements of stochasticity as opposed to Pix2Pix, which produces deterministic outputs.

As stated in Sec.~\ref{sec:data}, the mean age within each database differs significantly [$72\pm 14$ years (stroke) compared to $38\pm 24$ years (normal)]. Consequently, the stroke DB displays anatomical peculiarities that can be attributed to the advanced age of the patients, most notably enlarged ventricle volumes. For example, the mean lateral ventricle volumes are $(9.1\pm6.6)\,\mathrm{ml}$ in the normal DB versus $(20\pm12)\,\mathrm{ml}$ in the stroke DB. We have verified that all other anatomical features are consistent between the databases. Moreover, being mostly recorded on a 1.5\,T scanner, the stroke DB images have lower signal-to-noise ratios than those in the normal DB. To investigate whether this poses a systematic issue, we have generated synthetic DWIs from lesion-free segmentation masks in the stroke DB, but did not find any improvement, see Fig.~\ref{fig:experiments}.\footnote{This is also true if the ITM is trained only on stroke DB data and not all data combined.} Thus we conclude that synthetic data is beneficial if it introduces additional anatomical variation.

Finally, we have seen that all models perform well on large lesions, as can be seen in the upper panels of Fig.~\ref{fig:examples}. Conversely, the case shown in the bottom panels, which entails several smaller lesions, is not segmented equally well by all models; especially the U-Net trained only on clinical data produces false positive predictions. This appears to be a common feature and \emph{a posteriori} justifies the lower lesion threshold of 20 voxels, which inhibits too many false predictions. We have also trained models without any restriction on the lesion volume; however, found only worse performance and more false-positive labels.

Our study has several limitations that need to be addressed in future work. First, this is a single-center study using data from only one institution for training. While we did attempt to use the ISLES 2015 data for evaluating how well our models generalize, more quantitative analyses are necessary to guarantee robustness. Second, while DWIs are indeed the gold standard for lesion detection and quantification, CT is the clinically more common imaging modality, even though the lesion contrast is much lower. In the future, an important question will be how generative models can be used to generate more than one contrast. However, the issue of domain adaptation in medical imaging is subject of ongoing research, see e.g.~Refs.~\cite{intensityAug,billot2020learning,domainadaptation}.

Comparing our results to Ref.~\cite{FederauStroke}, two key aspects should be highlighted: There, the authors used masked lesions extracted from real DWIs of stroke patients to implant them into real DWIs of healthy patients, thus creating a similar database of synthetic images to train a segmentation network. While this creates fully realistic 3D images (at least outside the lesions), the approach is combinatorically limited, while our approach could in principle generate a continuum of new, unseen images either from a (continuous) latent space representation in the case of SPADE and the lesion generating GAN, or continuous deformations of segmentation masks. While there, results similar to ours were obtained using clinical data augmented with some 2000 synthetic cases, the authors did not discuss the use of synthetic images alone. A key result of our study is that a U-Net trained only on synthetic images generated by ITMs, cannot keep up with the results of a model trained on clinical data. However, such models can be fine-tuned on just a handful of clinical cases yielding result comparable to those obtained in a purely clinical setting with many high-quality segmentations. 

To address the question of 3D coherence, we have also attempted to train 3D ITMs. However, training 3D generative models is difficult, because these models tend to quickly exhaust available memory capacities such that the image resolution or the training batch size are limited. Moreover, GANs often do not converge when trained on 3D data.  Thus, it turns out that the proposed pipeline is the most economical setup, reliably producing  3D brain volumes and IS lesions of acceptable 3D coherence. This is owing to the fact that we used real anatomical segmentations to feed into the ITMs.

To put the results of this study into perspective, we report the results of a some recent approaches to the issue of stroke lesion segmentation. The aforementioned Ref.~\cite{FederauStroke} reported a maximal mean DSC of 72.1\% obtained on \emph{the same} test set as ours. This is to be compared with our maximal DSC of 70.8\%, in agreement within one standard deviation. Another approach comparable to ours was presented in Ref.~\cite{song2019generative} using a generative model to obtain DWIs from CT input images. This approach yielded a DSC of 62.4\% compared to a baseline DSC of 57.2\% without the generator model. This highlights possible applications of our method in cross-modality analyses, but also confirms our finding that the use of generative models generically enhances the quality of segmentations. Notice that the best DSC there compares to the DSC we obtain on clinical DWIs, but was obtained on CT, where IS is much less pronounced as is the case for DWIs. Next, Ref.~\cite{10.3389/fneur.2019.00541} chose to approach the segmentation of stroke lesions by using Markov random fields on FLAIR images obtaining a DSC of 58.2\%. Given the different contrast considered in this reference, this number cannot be directly compared to our results. However, the proposed method systematically outperforms segmentation approaches based on convolutional neural networks, thus providing a potential direction to investigate in the future. Similarly, Ref.~\cite{TOMITA2020102276} proposed a so-called zoom-in zoom-out data augmentation method that significantly improved the quality of neural network based segmentations of stroke lesions on T1-weighted MR images. This indicates that shape deformations could be a promising way to improve the quality of segmentations (also on synthetic images).

\section{Conclusions \& Outlook}\label{sec:conclusions}
We have investigated the applicability of DL-based image-to-image translation models to generalize data augmentation of medical image data sets. Our results highlight
that DL-based algorithms yield significantly better performance than traditional data augmentation techniques. This is in line with multiple recent studies, which have investigated the performance of different image-to-image translation models. However, only few publications 
exist to date, which subsequently study the potential improvement of DL-based segmentation and diagnosis systems, due to the synthetically enhanced training data sets~\cite{revisions1,revisions4}. Our work shows that this increase in data set size improves the quality of the segmentation network significantly.
In addition, it has a number of interesting consequences, such as using synthetic data to train disease detection algorithms, e.g. for IS lesion segmentation as in our case, 
avoids any data privacy obstructions. Thus, data in the form of trained ITMs can be made available to a much broader audience,which promises to accelerate the advances in machine 
learning applications in diagnostic medicine. Moreover, the technique can be used to augment available data sets with synthetic data.

Following up on our results, several directions are conceivable. As we have seen, training on synthetic data alone does not provide competitive results. A more careful data preparation could provide contrasts that are more aligned with those of the clinical data; however, at the price of fine-tuning the model to a single data set. Alternatively, using the idea of life-long learning~\cite{Thrun1998,TransferLearning}, a modified network architecture could be constructed, such as the one by~\cite{LifeLongMR}, which allows users to easily adapt the models to their data, even if acquired under entirely different circumstances. Incorporating more than one contrast for segmentation is another promising avenue. Finally, the proposed pipeline is straightforwardly generalized to other pathologies that can be identified on DWI, but also on other imaging modalities of the entire human body. A particularly interesting question is whether the diagnosis of rare diseases can be improved by augmenting available data sets by artificial data in order to compensate the intrinsic imbalance of the clinical training data. 

\section*{Acknowledgments}
The authors would like to thank Sebastian Kozerke and Thomas Joyce for useful discussions. 

\appendix \label{app:appendix}
\section{Model architectures}
In this appendix, we use the following abbreviations: Ck is a convolution, CTk a transposed convolution layer with k filters, D is a dropout layer, (L)ReLU is a (leaky) rectified linear unit activation layer, M a max-pooling layer, U a nearest neighbor upsampling layer, BN a batch normalization layer, and IN instance normalization.

\subsection{U-Net}
The 3D U-Net~\cite{RonnebergerUNet} consists of a downsampling and an upsampling branch, connected by a bottleneck layer. In the downsampling branch, we include 4 blocks D16-D32-D64-D-128, the upsampling path is U128-U64-U32-U16-C$n$, where $n$ is the number of segmentation labels, in this case $n=2$ (IS and non-IS), and the final convolution uses a softmax activation. The downsampling blocks D$k$ are consecutive layers C$k$-ReLU-D-C$k$-ReLU-D-M with stride-1 convolutions, while the upsampling blocks U$k$ are CT$k$-BN-ReLU-C$k$-ReLU-D-C$k$-ReLU-D with stride-1 convolutions and stride-2 transposed convolutions. To achieve the U-Net's characteristic skip connection, the output of each downsampling block is concatenated with the output of the transposed convolution in the upsampling branch at the matching image resolution. Between the two branches, the so-called bottleneck is a block C256-ReLU-D-C256-ReLU-D. The dropout rate is 2\% and  all convolutional kernels are $3\times3\times3$. The U-Net operates at an image resolution of $128\times128\times32$ and has a total of $5.6 \cdot 10^6$ trainable parameters. We train the model using a batch size of 5, reflecting memory restrictions.

\subsection{Pix2Pix}
The 2D Pix2Pix generator architecture is based on the U-Net, comprising a down- and an upsampling path. The details of the generator are largely identical to the proposed architecture by~\cite{pix2pix2017}. However, we replaced the batch-wise normalization with an instance-wise normalization~\cite{InstanceNorm}. The downsampling path is D64-D128-D256-D512-D512-D512-D512-D512, where the downsampling blocks  D$k$ are C$k$-IN-LReLU. Each convolutional layer contains a $4\times4$ kernel, stride 2, and is followed by the InstanceNorm. In the upsampling path we use upsampling blocks U256-U128-U64, each U$k$ consists of a sequence of layers $U$-C$k$-LReLU-D-IN-LReLU and stride-1 convolutions with kernel size $4\times4$. The downsampling block's output is concatenated before the final ReLU layer with a leakiness of 0.2. The generator model has a total of $41.8 \cdot 10^6$ trainable parameters, and we use a batch size of 8 for training.

\subsection{SPADE}
The SPADE generator samples a latent vector from a 128-dimensional random normal distribution centered at the origin with isotropic variance of unit magnitude. The latent vector is then passed to a dense layer with output size $4 \cdot 4 \cdot 128$ and reshaped into a feature map with dimension $4\times4\times128$. The following five residual blocks follow the original design by~\cite{SPADE}, including the SPADE normalization layers, which process the segmenattion maps. The design of each block is SPADE-LReLU-C$k$-SPADE-LReLU-C$k$, where the convolutions have a kernel size of $3\times3$ and unit stride as in the original setup; however, notice that we work with a reduced number of filters $k$ (128-64-32-16-8), the leaky ReLU layers have a slope of 0.2, and each SPADE block is followed by an upsampling layer until the desired resolution of $128\times128\times32$ is reached. A final $3\times3$ convolution is added at the end with a $\tanh$ activation and a single filter to match the data dimensions. The model is trained with a batch size of 8.

\subsubsection{Discriminator}
All ITMs use the PatchGAN discriminator~\cite{pix2pix2017,PatchGAN1, PatchGAN2}, a convolutional network whose output is not a single number as is the case for ordinary discriminator networks. Instead PatchGAN outputs an array of numbers, which, due to the convolutional character of the network, are connected only to a patch of the input image. The size and number of these patches depends on the number of convolutional layers. In the original Pix2Pix setup, where $256\times256$ images were considered, the receptive field was $70\times70$. Here we have chosen three convolutions, C32-IN-LReLU-C64-IN-LReLU-C128-IN-LReLU, which reduces the receptive field from $70\times70$ to $34\times34$ --~a choice which reflects the fact that the DWI slices under consideration measure $128\times128$. The final layer is followed by a stride-one, size-one filter, $4\times4$ convolution to map to the desired output shape. All other convolutions have kernel sizes $4\times4$, stride 2, and filters as indicated; and the leaky ReLU has a slope of 0.2.
The discriminator model has a total of 888,898 parameters.

\subsection{Fake lesion generation}
We have experimented with several architectures and loss functions and found that a Wasserstein GAN~\cite{Arjovsky2017WassersteinG} with a partial gradient loss~\cite{partialGPloss} yields the best results. This model can be trained on 3D lesion masks in a stable manner, and therefore provides the ideal building block for our synthesis pipeline. The output activation of this model is a softmax function, which we transform into a lesion prediction by thresholding. The GAN is trained on the 449 lesion masks that are available in the IS database with a batch size of 8 for 500 epochs. We use Instance normalization and Leaky ReLU (leakiness 0.2) activations after each convolution-upsampling step.

\bibliographystyle{IEEEtran}
\bibliography{literature}

\end{document}